# Labyrinthine domains in ferroelectric nanoparticles: a manifestation of gradient-induced morphological transition

*Eugene A. Eliseev[1], Yevhen M. Fomichov[1], Sergei V. Kalinin[2], Yulian M. Vysochanskii[3], Peter Maksymovich[2] and Anna N. Morozovska[4,5]\**

[1] *Institute for Problems of Materials Science, National Academy of Sciences of Ukraine, Krjijanovskogo 3, 03142 Kyiv, Ukraine*

[2] *The Center for Nanophase Materials Sciences, Oak Ridge National Laboratory, Oak Ridge, TN 37831*

[3] *Institute of Solid State Physics and Chemistry, Uzhgorod University, 88000 Uzhgorod, Ukraine*

[4] *Institute of Physics, National Academy of Sciences of Ukraine, 46, pr. Nauky, 03028 Kyiv, Ukraine*

[5] *Bogolyubov Institute for Theoretical Physics, National Academy of Sciences of Ukraine, 14-b Metrolohichna str. 03680 Kyiv, Ukraine*

## Abstract

In the framework of the Landau-Ginzburg-Devonshire (LGD) approach we studied finite size effects of the phase diagram and domain structure evolution in spherical nanoparticles of uniaxial ferroelectric. The particle surface is covered by a layer of screening charge characterized by finite screening length. The phase diagram, calculated in coordinates "particle radius – screening length" has a wide region of versatile poly-domain structures separating single-domain ferroelectric and nonpolar paraelectric phases. Unexpectedly, we revealed a region of stable irregular labyrinthine domains in the nanoparticles of uniaxial ferroelectric $CuInP_2S_6$ with the first order paraelectric-ferroelectric phase transition. We established that the origin of labyrinthine domains is the mutual balance of LGD, polarization gradient and electrostatic energies. The branching of the domain walls appears and increases rapidly when the polarization gradient energy decreases below the critical value. Allowing for the generality of LGD approach, we expect that the gradient-induced morphological transition can be the source of labyrinthine domains appearance in many spatially-confined ferroics with long-range order parameter, including relaxors, ferromagnetics, antiferrodistortive materials and materials with incommensurate ferroic phases.

---

\* corresponding author, e-mail: anna.n.morozovska@gmail.com

# I. INTRODUCTION

The ferroic materials described by Landau theory of symmetry-breaking have a substantial impact on fundamental science and various applications. Different types of topological defects in different ferroics (ferromagnets, ferroelecrics, ferroelastics) are even more numerous and enigmatic than different types of symmetry-breaking, and consequently, they become one of the key fundamental problems and hot topics in scientific community [1, 2].

Complementary to the topological point defects [1], domain walls can be considered as extended 2D topological defects in ferroics (see e.g. chapter 8 in [2] and refs. therein). Vortices and vertices composed by the closure of four domain walls have been observed experimentally and described theoretically in a bulk and nanosized ferroelectrics [3, 4, 5, 6, 7]. Stable surface-induced labyrinthine domain structures were observed by Piezoresponse Force Microscopy (PFM) in ergodic ferroelectrics relaxors and explained by the presence of higher-order term in free-energy expansion that gives rise to the polarization modulations [8]. Fractal domain structures are sometimes observed in multiferroic thin films [9] and near the surface of relaxors close to relaxor-ferroelectric boundary [10], but the labyrinthine domains with a single characteristic length scale were observed by PFM in ergodic relaxors only [8]. These labyrinthine structures can coexist with classical ferroelectric domains closer to ferroelectric composition limit [11, 12]. The labyrinthine domain structure was predicted theoretically in thin films of incommensurate ferroelectrics [13] and bi-layered ferroelectrics [14], being similar to those observed in ultrathin magnetic films [15].

However, we did not find any experimental observation or theoretical prediction of labyrinthine domains in the nanoparticles of ordered ferroelectrics, which intriguing polar and dielectric properties attract permanent attention of researchers. Classical examples are unexpected experimental results of Yadlovker and Berger [16, 17, 18], which reveal the enhancement of polar properties of cylindrical nanoparticles of Rochelle salt. Frey and Payne [19], Zhao et al [20], Drobnich et al [21], Erdem et al [22] and Golovina et al [23, 24, 25] demonstrate the possibility to control the phase transitions (including new polar phases appearance) for $BaTiO_3$, $S_2P_2S_6$, $PbTiO_3$ and $KTa_{1-x}Nb_xO_3$ nanopowders and nanoceramics by **finite size effects**.

The theory of finite size effects in nanoparticles allows one to establish the physical origin of the polar and dielectric properties anomalies, and phase diagrams changes appeared under the nanoparticles sizes decrease. In particular, using the continual phenomenological approach Niepce [26], Huang et al [27, 28], Ma [29], Eliseev et al [30] and Morozovska et al [31, 32, 33] have shown, that the changes of the transition temperatures, the enhancement or weakening of polar properties in a **single-domain** spherical and cylindrical nanoparticles are conditioned by the various physical mechanisms, such as correlation effect, depolarization field, flexoelectricity, electrostriction, surface tension and Vegard-type chemical pressure.

Notably depolarization field always decreases ferroelectric polarization and transition temperature, especially under the presence of imperfect screening [34, 35, 36]. For majority of models the particles were regarded covered with perfect electrodes and so their single-domain state would be stable. Only few models describing the imperfect screening effect in nanoparticles have been evolved [34-36].

To fill the gap in the knowledge, below we analyze the phase diagram and domain structure evolution in spherical nanoparticles of uniaxial ferroelectric CuInP$_2$S$_6$ (**CIPS**). We regarded that the particle surface is covered by a layer of screening charge characterized by finite screening length. The imperfect screening and finite size effects are studied using the Landau-Ginzburg-Devonshire (**LGD**) approach combined with the electrostatic equations. We revealed that a regular stripe domain structure avalanche-like transforms into a labyrinth pattern with a gradient term decrease below the critical value and classified the event as a **gradient-induced** morphological transition.

The applicability of LGD approach for thin films and nanoparticles with radius less than (2-10) nm is corroborated by the fact, that the critical sizes of the long-range order appearance and properties calculated from atomistic [37, 38, 39, 40, 41] and phenomenological [30-33, 42, 43] theories are in a good agreement with each other as well as with experimental results for nanosized ferromagnetics [44] and ferroelectrics [16-20, 22, 45]. Both atomistic simulations and LGD-description are absent for CIPS nanoparticles.

## II. THEORETICAL APPROACH

Let us consider a CIPS nanoparticle of radius $R$ with a one-component ferroelectric polarization $P_3(\mathbf{r})$ directed along the crystallographic axis 3 [**Fig.1(a)**]. The particles are covered by a layer of screening charge with a surface charge density $\sigma$ characterized by a nonzero screening length $\lambda$. The specific nature of the surface charge can be, e.g., Bardeen-type surface states [46]. For the case the screening charges can be localized at surface states caused by the strong band-bending via depolarization field [47, 48, 49, 50, 51], at that $\lambda$ can be much smaller (≤0.1 nm) than a lattice constant (~0.5 nm) [34]. Concrete expression for $\lambda$ can be derived in, e.g., Stephenson-Highland ionic adsorption model [52, 53, 54], by the linearization of $\sigma$, as $\sigma \approx -\varepsilon_0 \varphi/\lambda$, where $\lambda^{-1} \approx \sum_i \frac{(eZ_i)^2}{4\varepsilon_0 A_i k_B T}\left(1 - \tanh^2\left(\frac{\Delta G_i^{00}}{2 k_B T}\right)\right)$, $Z_i$ is the ionization number of the surface ions, $1/A_i$ is their saturation densities, $\Delta G_i^{00}$ is the free energy of the surface ion formation, $\varepsilon_0$ is a universal dielectric constant. In general case $\lambda$ depends on temperature $T$ and screening charges nature. Since we do not know the temperature dependence of $\lambda$, we performed calculations regarding $\lambda$ changing in the range $(10^{-3} - 1)$ nm.

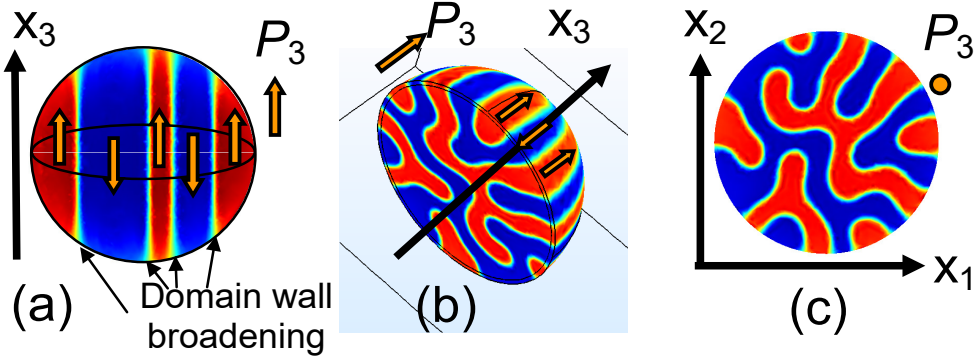

**FIG. 1.** Labyrinthine domains in a spherical CIPS nanoparticle. **(a)** Polar cross-section, **(b)** semi-spherical view and **(c)** equatorial cross-section. Radius $R$=10 nm, screening length $\lambda$=0.03 nm, room temperature 293 K. CIPS parameters are listed in **Table I**.

For a layered perovskite with layers plane (001) and ferroelectric dipoles directed in the out-of-plane direction, we can assume that the dependence of the in-plane components of electric polarization on the inner field electric $E_i$ is linear $P_i = \varepsilon_0(\varepsilon_b - 1)E_i$ ($i$ = 1, 2), where an isotropic background permittivity $\varepsilon_b$ is relatively small, $\varepsilon_b \leq 10$ [55]. Polarization component $P_3(\mathbf{r})$ contains background and soft mode contributions. Electric displacement vector has the form $\mathbf{D} = \varepsilon_0 \varepsilon_b \mathbf{E} + \mathbf{P}$ inside the particle and $\mathbf{D} = \varepsilon_0 \varepsilon_e \mathbf{E}$ outside it; $\varepsilon_e$ is the relative dielectric permittivity of external media regarded unity (air or vacuum).

Euler-Lagrange equation for the ferroelectric polarization $P_3(\mathbf{r})$ follows from the minimization of LGD free energy functional $G = G_{Landau} + G_{grad} + G_{el} + G_{es+flexo}$, that includes Landau expansion, $G_{Landau}$, polarization gradient energy contribution, $G_{grad}$, electrostatic contribution $G_{el}$, and elastic, electrostriction and flexoelectric contributions $G_{es+flexo}$ (see e.g. [30, 35, 56]):

$$G_{Landau} = \int_{|\vec{r}|<R} d^3r \left( \frac{\alpha}{2} P_3^2 + \frac{\beta}{4} P_3^4 + \frac{\gamma}{6} P_3^6 \right), \tag{1a}$$

$$G_{grad} = \int_{|\vec{r}|<R} d^3r \left( \frac{g_{11}}{2}\left(\frac{\partial P_3}{\partial x_3}\right)^2 + \frac{g_{44}}{2}\left[\left(\frac{\partial P_3}{\partial x_2}\right)^2 + \left(\frac{\partial P_3}{\partial x_1}\right)^2\right] \right), \tag{1b}$$

$$G_{el} = -\int_{|\vec{r}|<R} d^3r \left( P_3 E_3 + \frac{\varepsilon_0 \varepsilon_b}{2} E_i E_i \right) - \int_{|\vec{r}|=R} d^2r \left( \frac{\varepsilon_0 \varphi^2}{2\lambda} \right) - \frac{\varepsilon_0 \varepsilon_e}{2} \int_{|\vec{r}|>R} E_i E_i d^3r, \tag{1c}$$

$$G_{es+flexo} = \int_{|\vec{r}|<R} d^3r \left( -\frac{s_{ijkl}}{2}\sigma_{ij}\sigma_{kl} - Q_{ij3}\sigma_{ij}P_3^2 - F_{ijk3}\left(\sigma_{ij}\frac{\partial P_3}{\partial x_k} - P_3 \frac{\partial \sigma_{ij}}{\partial x_k}\right) \right). \tag{1d}$$

Here $E_i$ are electric field components related with electric potential φ as $E_i = -\partial\varphi/\partial x_i$. The coefficient α linearly depends on temperature $T$, $\alpha = \alpha_T(T - T_C)$, where $T_C$ is the Curie temperature. The coefficient β is temperature-independent and negative, since CIPS undergoes the first order transition to paraelectric phase. Coefficient γ and gradient coefficients $g_{11}$ and $g_{44}$ are positive and temperature independent. An isotropic approximation, $g_{44} \approx g_{55}$ in (001) plane was taken for monoclinic CIPS structure. $\sigma_{ij}$ is the stress tensor in Eq.(1d). We omit the evident form of the $G_{es+flexo}$ for the sake of simplicity, it is listed in Refs.[57, 58, 59]. Since the values of the electrostriction and flexoelectric tensor components, $Q_{ijkl}$ and $F_{ijkl}$, are unknown for CIPS, we performed numerical calculations with the coefficients varied in a physically reasonable range ($|F_{ijkl}| \leq 10^{11}$ m$^3$/C and $|Q_{ijkl}| \leq 0.1$ m$^4$/C$^2$). Results proved the insignificant impact of electrostriction and flexoelectric coupling on domain morphology [60]. Other LGD parameters for a bulk ferroelectric CIPS were taken from Ref.[61] and are listed in **Table I**.

**Table I.** LGD parameters for bulk ferroelectric CuInP$_2$S$_6$ used in calculations

| $\varepsilon_b$ | $\alpha_T$(C$^{-2}$·m J/K) | $T_C$ (K) | β (C$^{-4}$·m$^5$J) | γ (C$^{-6}$·m$^9$J) | $g_{11}$ (m$^3$/F) [62] | $g_{44}$ (m$^3$/F) |
|---|---|---|---|---|---|---|
| 7 | 1.569×10$^7$ | 302 | −1.8×10$^{12}$ | 2.2×10$^{15}$ | 1.0×10$^{-10}$ | vary in the range(0.3 – 3)×10$^{-11}$ |

Corresponding Euler-Lagrange equation for $P(\mathbf{r}_3)$ has the form:

$$\alpha(T)P_3 + \beta P_3^3 + \gamma P_3^5 - g_{44}\left(\frac{\partial^2}{\partial x_1^2} + \frac{\partial^2}{\partial x_2^2}\right)P_3 - g_{11}\frac{\partial^2 P_3}{\partial x_3^2} = E_3. \quad (2)$$

The boundary condition for $P$ at the spherical surface is natural, $\partial\vec{P}/\partial\mathbf{n}\big|_{r=R} = 0$, **n** is the outer normal to the surface. The potential φ satisfies a Poisson equation inside the particle,

$$\varepsilon_0\varepsilon_b\Delta\varphi = -\frac{\partial P}{\partial x_3}, \quad (3a)$$

and Laplace equation outside it,

$$\Delta\varphi = 0. \quad (3b)$$

3-D Laplace operator is denoted by the symbol $\Delta$. Equations (3) are supplemented by the condition of potential continuity at the particle surface, $(\varphi_{ext} - \varphi_{int})\big|_{r=R} = 0$. The boundary condition for the normal components of electric displacements, $(\mathbf{n}(\mathbf{D}_{ext} - \mathbf{D}_{int}) + \sigma)\big|_{r=R} = 0$, where the surface charge density $\sigma = -\varepsilon_0\varphi/\lambda$.

# III. NUMERICAL RESULTS AND DISCUSSION

## A. Main features of phase diagrams at different temperatures

Phase diagrams was studied at different temperatures in the range (300 – 200) K. Unfortunately we do not know the temperature dependence of λ, and so we perform all calculations regarding λ changing in the range ($10^{-3}$ – 1) nm. Phase diagram of CIPS nanoparticles calculated at T=293 K and 200 K in coordinates "radius *R* – screening length λ" is shown in **Fig. 2(a)** and **2(b),** respectively**.**

At room temperature the phase diagram has an unexpectedly wide region of stable poly-domain states (**PDFE**) separating single-domain ferroelectric (**SDFE**) and nonpolar paraelectric (**PE**) phases [see **Fig. 2(a)**]. The bottom row shows the typical changes of polarization distribution in the equatorial cross-section of the nanoparticle with *R*=5 nm, which happens with increase of λ. A single-domain state is stable at very small λ<0.01 nm**,** two-domain structure (electric quadrupole) is stable in the interval 0.01<λ<0.017 nm, three-domain structure (electric octupole) exist at 0.017<λ<0.019 nm, 2N-multipolar domain stripes are stable at 0.02<λ<0.035 nm. Coexistence of PDFE and PE phase when the nanoparticle consists of PE surface layer and ferroelectric domain stripes in the core appears at 0.035<λ<0.045 nm, and is followed by the size-induced phase transition into a stable PE at λ>0.045 nm. Unexpectedly, we revealed a region of stable "labyrinthine" domains (**SLD**) of irregular shape (yellow circles) inside the region of regular domain structures with quadruple two (purple circles) or multiple (magenta circles) domain stripes. SLD region is within a dashed parallelogram in **Fig.2(a).**

With the temperature decrease from 293 K (that is very close to the CIPS Curie temperature 302 K) to 200 K the region of a SDFE significantly increases towards smaller radii *R* (up to the very small *R*=1 nm for which LGD applicability becomes questionable) and higher λ (from e.g. λ=0.003 nm at 293 K to 0.02 nm at 200 K) [compare the size of SDFE regions in **Fig.2(a)** and **2(b)**]. The wide region of PE phase (present at 293 K) almost disappears with the temperature decreasing to 200 K leading to the conclusion that PDFE state can be stable in ultra-small CIPS nanoparticles (with radius less than 2 nm) covered by a screening charge [compare the size of PE regions in **Fig.2(a)** and **2(b)**]. The shift and increase of stable labyrinthine domains region(s) are evident with the temperature decrease from 300 to 200 K [compare the size and positions of SLD regions in **Fig.2(a)** and **2(b)**]. The increase of SDFE, PDFE and SLD regions with the temperature *T* decrease stem from the well-established fact that FE phase becomes deeper and wider with the temperature increase, since the coefficient $\alpha = \alpha_T(T - T_C)$ acquires higher negative values with *T* decrease below Curie temperature $T_C$.

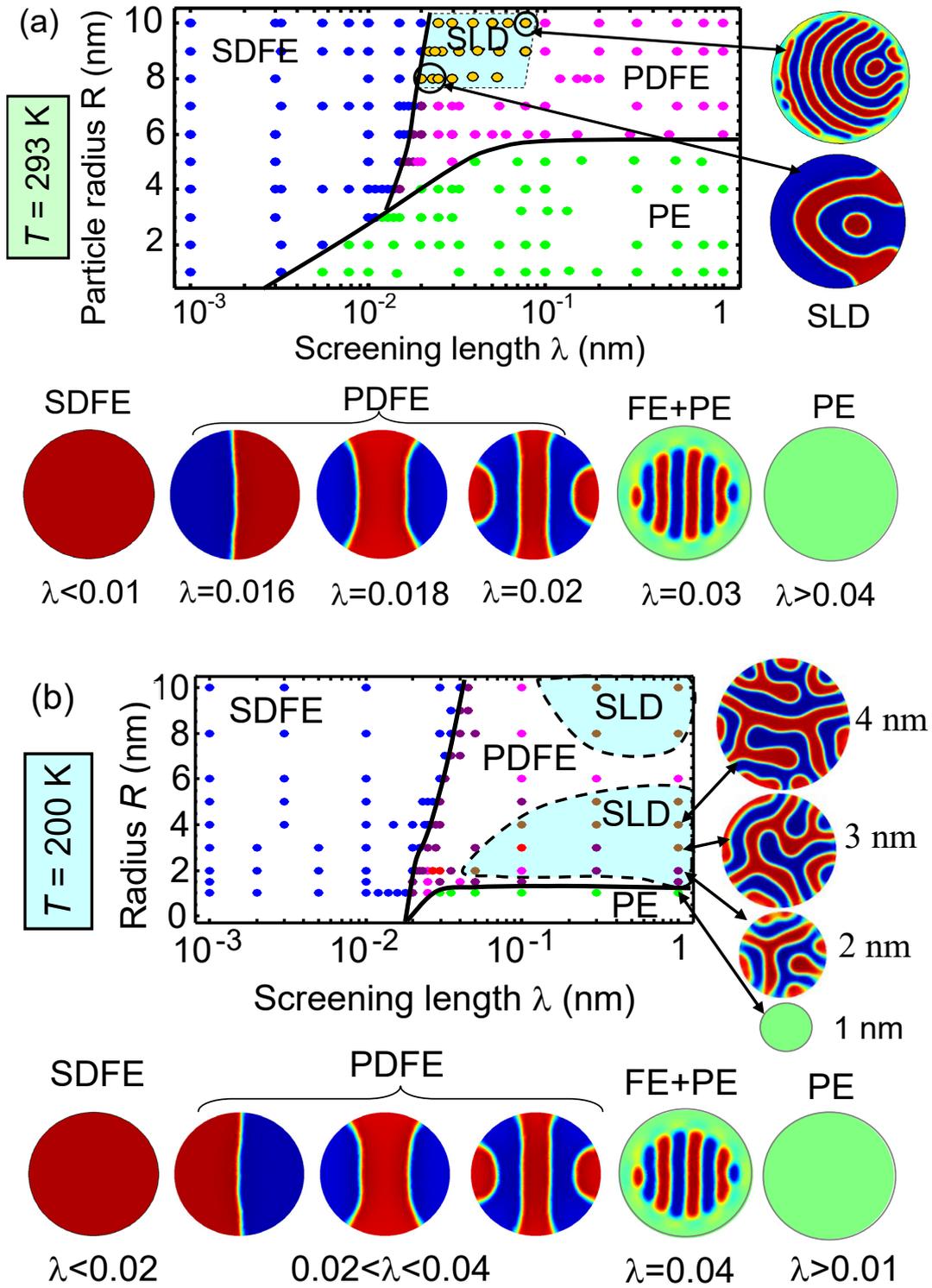

**FIG. 2.** Phase diagram of CIPS nanoparticles in coordinates "radius $R$ – screening length $\lambda$" calculated for the gradient coefficient $g_{44}=2\times10^{-11}$ m$^3$/F and temperatures 293 K **(a)** and 200 K **(b)**. The ferroelectric single domain (SDFE), ferroelectric poly domain (PDFE) and paraelectric (PE) phases are stable. The region of stable labyrinthine domains (SLD) is located within dashed light-blue regions. The bottom rows shows typical polarization distributions in the equatorial cross-sections of the nanoparticles with radius $R$=5 nm and different values of $\lambda$ (in nm). CIPS parameters are listed in **Table I.**

The effect of geometric catastrophe can be imagined from the images of SLD in the nanoparticles of radius 4, 3 and 2 nm for which the number of branches and sharp bendings of domain walls gradually decreases with the particle radius decrease from 4 nm to 2 nm [see right column in **Fig. 2(b)**]. Eventually SLD disappears for $R$=1 nm. Hence the effect of geometric catastrophe suppresses the SLD in small particles.

Note the validity of our prediction regarding SLD appearance and PDFE state conservation for nanoparticles of sizes more than $2R = 4$ nm, because they corresponds to 10 lattice constants or more. It is general opinion that LGD approach can be valid only qualitatively for the sizes less than 10 *l.c.* [16-20, 22, 30-33], and must be approved by *ab initio* calculations.

**B. Labyrinthine domains stability and evolution**

Polarization distributions in the equatorial cross-section of the nanoparticle with radius $R$=10 nm, screening length λ=0.03 nm and room temperature are shown in **Fig. 3.** The energy values computed for the single-domain [**Fig.3(a)**], two-domain [**Fig.3(b)**], three-domain [**Fig.3(c)**], eight-domain stripes [**Fig.3(d)**] and the most stable labyrinthine domain structure [**Fig.3(e)**] are $G$ = –5.98, –6.94, –7.08, –8.55 and –9.04 (in $10^{-20}$ J) at fixed value of $g_{44}$=2×$10^{-11}$ m$^3$/F. Thus the "labyrinthine" structure has the minimal energy corresponding the optimal balance between the gradient-correlation energy (1b) tending to minimize the area of the domain walls (and hence to decrease the number of them) and electrostatic energy (1c) decreasing with domain width decrease. Note that the walls of SLD are uncharged in the central part of the particle and become charged and broadened near its poles (see yellow-blue regions near the poles in **Figs. 1(a)** and **Fig. S2** in Ref.[60]), since their broadening causes the depolarization field decrease [63].

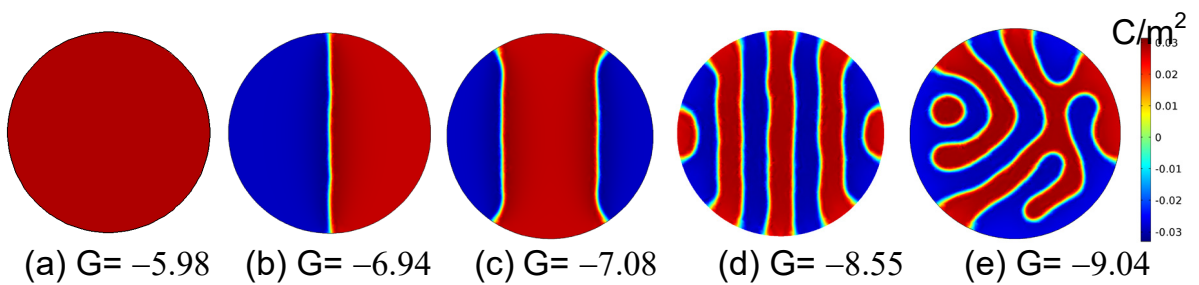

(a) G= –5.98    (b) G= –6.94    (c) G= –7.08    (d) G= –8.55    (e) G= –9.04

**FIG. 3.** Polarization distributions in the equatorial cross-section of the nanoparticle with $R$=10 nm, λ=0.03 nm, $g_{44}$=2×$10^{-11}$ m$^3$/F and room temperature 293 K. Cross-sections **(a)-(e)** correspond to different morphologies of the domain structure, namely single-domain state **(a),** two-domain **(b),** three-domain **(c),** multiple stripe domains **(d),** and labyrinthine domains **(e)**. The scale bar is for polarization $P_3$ in C/m$^2$. Values of the free energy G are listed below in $10^{-20}$ J. CIPS parameters are listed in **Table I**.

An example of labyrinthine domains evolution with increase of the gradient coefficient $g_{44}$ is illustrated in **Fig.4(a)-(e)**. Stable labyrinths exist at $g_{44}$ less than the critical value $g_{cr} \approx 2.5 \times 10^{-11}$ m$^3$/F [**Fig. 4(a)-(e)**], then they transforms into quasi-regular domain stripes [**Fig. 4(f)**], which in turn disappears with $g_{44}$ increase more than $2.8 \times 10^{-11}$ m$^3$/F [**Fig. 4(g)**]. Complementary we made sampling over 5 – 20 different labyrinthine domain patterns for each $g_{44}$, $R$ and $\lambda$ values, which emerged with computation time from different initial random distributions of polarization inside the particle (see **Fig. S1** in Ref.[60]). From **Fig.S1** we concluded the branching number $\Sigma$, defined as the total number of branched domain walls, dangling branches, separated stripes and loops, decreases sharply with $g_{44}$ increase. $\Sigma$ varies slightly for different samples far from the critical value $g_{cr}$ (top lines in **Fig.S1**), but the variation becomes bigger near the critical value $g_{cr}$ (bottom lines in **Fig.S1**).

Examples of $\Sigma$ calculation are shown in **Figs.5**. A color image of complex labyrinthine pattern with dangling branches, branch seedings, separated island and separated curved stripe is shown in **Fig.5(a)**. **Fig.5(b)** shows the black domains with white walls corresponding to the structure **(a)**. Graphs **(c, d)** with numbered features have been drawn allowing for the connectivity between different domains and particle surface in plot **5(b)**. Proposed algorithm of $\Sigma$ calculation counts all branching points, dangling branches and separated stripes ends, which do not cross the particle from one surface to another one. Meanwhile the straight or slightly curved stripes (even very small) that cross the particle from one surface to another one do not contribute to $\Sigma$. However, the algorithm is not ideal, because the criteria distinguishing "slightly curved" and "strongly curved" stripes are somehow voluntary. Actually, for some complex cases, like the one shown in **Fig.5(a)**, visual re-calculation of $\Sigma$ lead to different results for peculiarities number corresponding to "red" ($\Sigma_r=7$) and "blue" ($\Sigma_b=9$) domains [compare **Fig. 5(b)** and **5(c)**]. To improve the situation we operate with the values averaged for "red" and "blue" domains, e.g. $\Sigma=8$ corresponds to **Fig.5(a)**.

Graph **Fig.5 (e)** has no relation to plot **(a),** but it is characteristic for the simpler domain patterns close the transition to SLD, where the accuracy in $\Sigma$ calculation is the most important to establish the critical value $g_{cr}$ correctly [compare **Fig.5 (e)** with **Fig.4 (e)**]. **Fig.5(e)** illustrates how the branching point (number 5), dangling branches (numbers 1 and 2), and separated stripe ends (numbers 3 and 4), which do not cross the particle from one surface to another one, contribute to $\Sigma$ number. Other four slightly curved stripes, which both ends at particle surfaces are marked with asterisks "*", do not contribute to $\Sigma$ number.

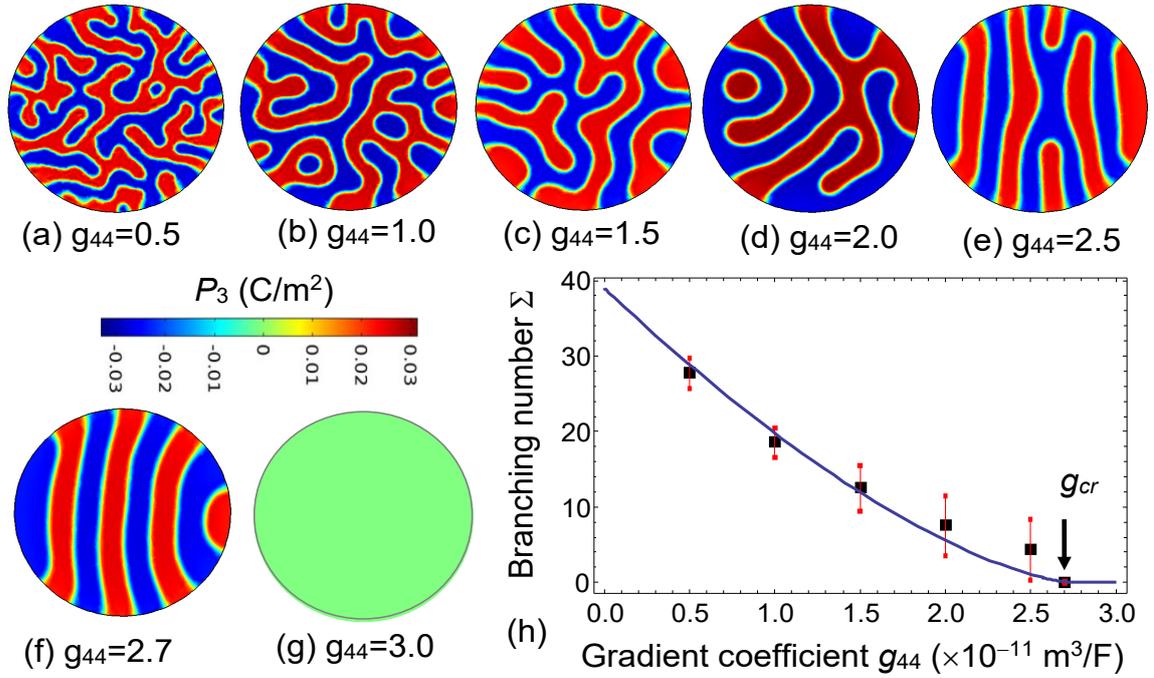

**FIG.4.** Evolution of labyrinthine domain structure in a CIPS nanopariticle with increase of the gradient coefficient $g_{44}$ (in $10^{-11}$ m$^3$/F) [plots **(a)-(g)**] for the screening length $\lambda$=0.03 nm, radius R=10 nm and room temperature 293 K. The scale bar is polarization value in C/m$^2$. **(h)** Dependence of the SLD branching number $\Sigma$ on $g_{44}$. Error bars corresponds to different samples of SLD emerging from different initial seeding. Black circles are averaged value $\langle\Sigma\rangle$ approximated by the function $\langle\Sigma\rangle = 39(1 - g_{44}/g_{cr})^{3/2}$ with $g_{cr} = 2.75\times10^{-11}$ m$^3$/F (blue curve). CIPS parameters are listed in **Table I**.

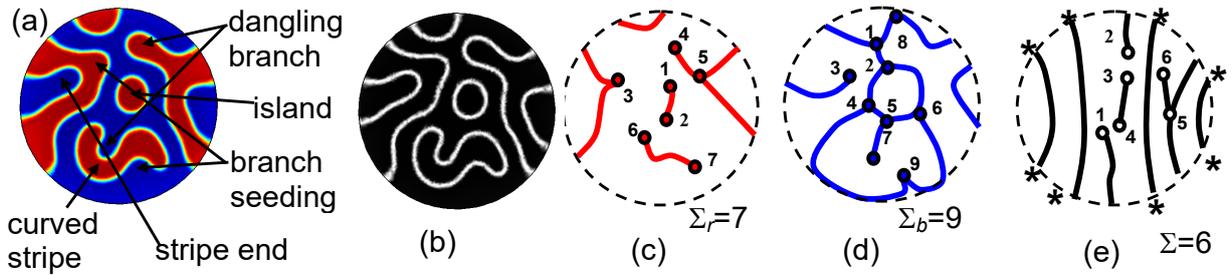

**FIG. 5. Examples of $\Sigma$ calculation using graph method. (a)** A color image of complex labyrinthine pattern with dangling branch, branch seeding, separated island and separated curved stripe. **(b)** Black domains with white walls corresponding to the structure **(a)**. Graphs **(c, d)** with numbered features have been drawn allowing for the connectivity between different domains and particle surface in plot **(b)**. Graph **(e)** has no relation to plot **(a)**, but it is characteristic for the patterns near the transition to SLD [compare **Fig.5 (e)** with **Fig.4 (e)**].

The sampling-averaged value $\langle\Sigma\rangle$ is not integer for fixed $g_{44}$. From **Fig.4(h),** the dependence of $\langle\Sigma\rangle$ on the gradient term $g_{44}$ is described by the function $\langle\Sigma\rangle = 32(1 - g_{44}/g_{cr})^{3/2}$, and so it continuously appears at $g_{44} = g_{cr}$. Hence we can associate $\langle\Sigma\rangle$ appearance with a rapid change of the domain walls connectivity.

We leave for further studies answer on the question how the threshold of labyrinthine domain appearance and the ranges of their stability at phase diagram can be derived analytically.

The theoretical prediction of labyrinthine domains requires urgent verification by PFM, that is an ideal tool for 3D visualization of the domain structure with nanoscale resolution (see e.g. [8, 64, 65, 66, 67] and refs therein). We are convinced by a numerical calculation that qualitatively similar SLD can be realized in other incompletely screened uniaxial ferroelectric nanoparticles, such as $Sn_2P_2(S,Se)_6$ and $LiNbO_3$, with the sizes near the first-order PE-FE transition. Notably the phase diagrams in **Figs.2** change drastically for $\beta \geq 0$ corresponding to the second order PE-FE transition, and we see neither PDFE-PE coexistence nor stable LD. Much more complex situation (corresponding to the balance of labyrinthine domains in the bulk and vortices at the surface) are expected in multiaxial ferroelectric nanoparticles with polarization rotation allowed, such as $BaTiO_3$, $BiFeO_3$, however we leave a discussion of these results for further studies.

## IV. CONCLUSION

In the framework of LGD approach combined with the equations of electrostatics, we studied the finite size effects of the phase diagrams and domain structure in spherical ferroelectric nanoparticles covered by a layer of a screening charge with finite screening length. The phase diagrams, calculated in coordinates "particle radius – screening length", has a wide region of versatile poly-domain states separating single-domain ferroelectric and nonpolar paraelectric phases. Quite unexpectedly we revealed that a regular stripe domain structure sharply transforms into a labyrinth pattern with a gradient term decrease below the critical value and named the event as a gradient-driven transition. Obtained results calculated for $CuInP_2S_6$ can be readily generalized for other incompletely screened nanoparticles of uniaxial ferroelectrics with the first order transition to the paraelectric phase.

**Acknowledgements.** S.V.K. and P.M. study was supported by the U.S. DOE, Office of Basic Energy Sciences (BES), Materials Sciences and Engineering Division (MSED) under FWP Grant No DEAC0500OR22725. A portion of this research was conducted at the Center for Nanophase Materials Sciences, which is a DOE Office of Science User Facility. A.N.M. work was partially supported by the National Academy of Sciences of Ukraine (project No. 0118U003375 and No. 0117U002612) and by the Program of Fundamental Research of the Department of Physics and Astronomy of the National Academy of Sciences of Ukraine (project No. 0117U000240).

**Authors' contribution.** E.A.E. wrote the codes, performed numerical calculations and prepared figures. Y.M.F. tested the codes and assisted E.A.E. with simulations. A.N.M. generated research idea, stated the problem, interpreted results and wrote the manuscript. S.V.K., Y.M.V. and P.M. worked on the results discussion and manuscript improvement.

# SUPPLEMENT TO THE MANUSCRIPT
# "Labyrinthine domains in ferroelectric nanoparticles: a manifestation of gradient-induced morphological transition"

## S.1. Impact of the electrostrictive and flexoelectric coupling

The codes for calculation of polarization, electric and elastic field distributions inside and outside the nanoparticle included both electrostrictive and flexoelectric coupling (see e.g. our previous publications for spherical nanoparticles [1] and thin films [2, 3, 4]). Corresponding contributions of electrostrictive and flexoelectric coupling to the free energy has the form

$$G_{es+flexo} = \int_{|\vec{r}|<R} d^3r \left( -\frac{s_{ijkl}}{2} \sigma_{ij}\sigma_{kl} - Q_{ijkl}\sigma_{ij}P_k P_l - F_{ijkl}\left( \sigma_{ij}\frac{\partial P_l}{\partial x_k} - P_l\frac{\partial \sigma_{ij}}{\partial x_k} \right) \right), \qquad (1d)$$

where $\sigma_{ij}$ is the stress tensor, electrostriction and flexoelectric tensor components are $Q_{ijkl}$ and $F_{ijkl}$, respectively. Also equation of mechanical equilibrium should be solved with elastic boundary conditions. Electric boundary conditions should be modified by flexoeffect as explained in details in Refs. [2-4]. To the best of our knowledge the values of the electrostriction and flexoelectric tensor components are unknown for CIPS.

Our previous results [2-4] for PbTiO$_3$ and BaTiO$_3$ thin films with all known coefficients and preliminary numerical calculations made for nanoparticles in this work with the coefficients varied in a physically reasonable range (fleaxoelectric tensor $|F_{ijkl}| \leq 10^{11}$ m$^3$/C and electrostriction tensor $|Q_{ijkl}| \leq 0.1$ m$^4$/C$^2$) proved that that the impact of electrostriction leads to no more that 1% strain of the particle as well as to slightly visible decrease of the domain wall width. Hence the change of the nanoparticle spherical shape to ellipsoidal one (it indeed reveals the tiny elongation or contraction depending on the electrostriction coefficients sign in polar direction z) leads to rather small changes of electric field, which do not affect significantly on the morphology of domain structure, both labyrinthine and regular ones.

The flexoelectric impact is really small, but can be non-trivial for uniaxial and multiaxial ferroelectrics, since it can charge the nominally uncharged Ising-type walls [5]. However for the case of CIPS we did not see any significant changes of the domain morphology or walls charging with the flexoefffect included.

## S.2. Sampling over different labyrinthine domain patterns

We made sampling over 10 – 20 different labyrinthine domain patterns, which emerged with computation time from different initial random distributions of polarization inside the particle (further named "initial seedings") for each $g_{44}$, $R$, $\lambda$ and $T$ values. Six labyrinthine domain patterns corresponding to definite $g_{44}$ values, which varies from $0.5\times10^{-11}$ m³/F (top row) to $2.5\times10^{-11}$ m³/F (bottom row), temperature $T$=300 K, particle radius $R$=10 nm, screening length $\lambda$=0.03 nm are shown in **Fig. S1**. The number of branches $\Sigma$ of the labyrinthine domain structure is indicated below each pattern in the figure, and it is different for different "samples", while the width of the labyrinth pattern is independent on the sample (it depends on $g_{44}$, $R$, $\lambda$ and $T$ and material parameters of CIPS). $\Sigma$ varies slightly for different samples far from the critical value $g_{cr}$ (see top lines in **Fig.S1**), but the variation becomes bigger near the critical value $g_{cr}$ (see bottom lines in **Fig.S1**). Naturally, the average value of $\Sigma$ calculated to the definite $g_{44}$ is not integer.

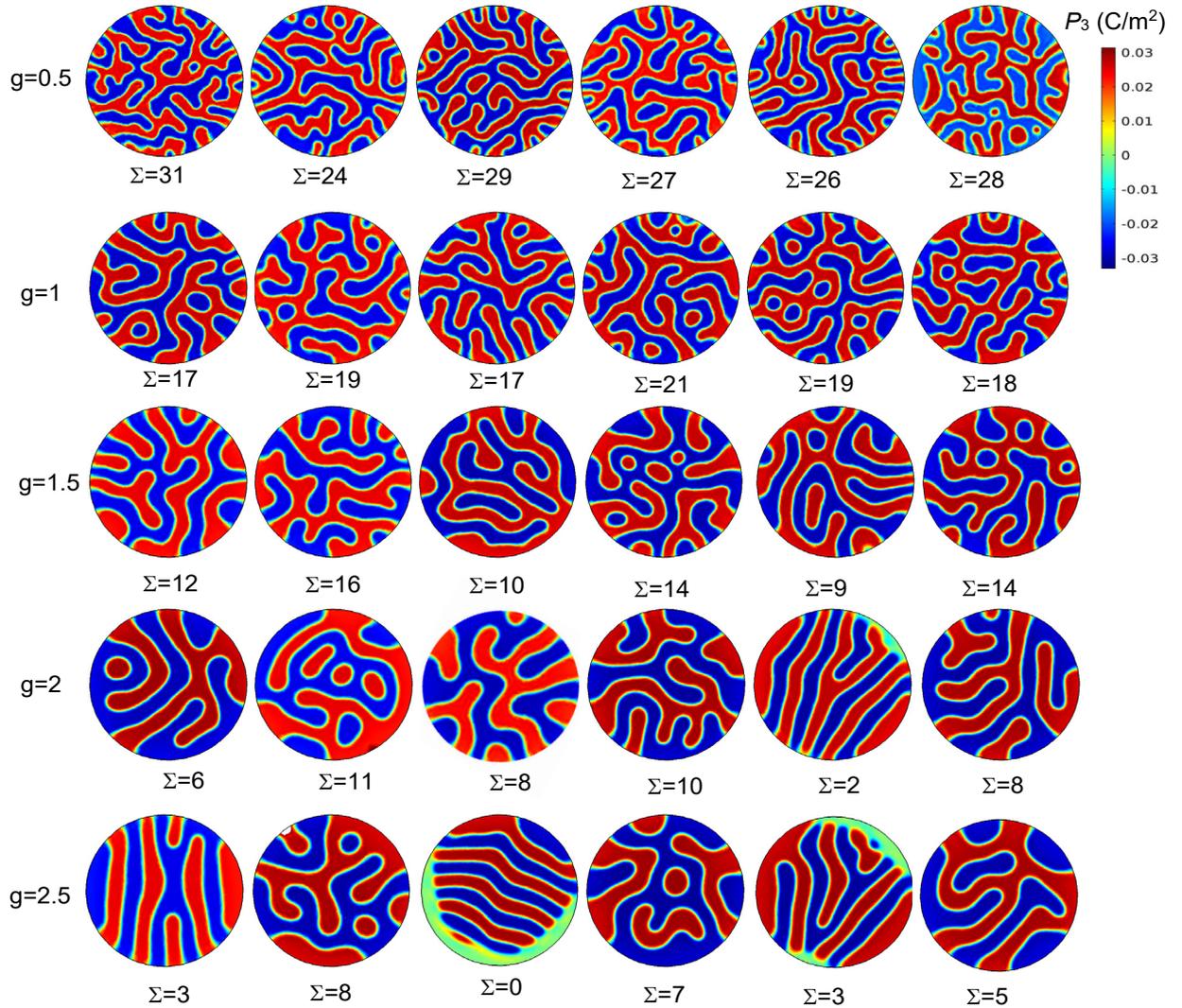

**FIG. S1.** Different samples of labyrinthine domains emerging with time from different random distributions of polarization in CIPS nanoparticles. The gradient coefficient $g_{44}$ varies from $0.5\times10^{-11}$ m³/F (top row) to $2.5\times10^{-11}$ m³/F (bottom row), temperature T=300 K, particle radius R=10 nm, screening length $\lambda$=0.03 nm. Other parameters are listed in **Table I**. The number of branches $\Sigma$ of the labyrinthine domain structure is indicated below each figure. The algorithm of Σ calculation is illustrated in **Fig. 5** and discussed in the main text**.**

Also we used different random seeding to generate regular domain stripes, and, as anticipated for the case, the results appeared independent of the initial seeding. We compared the energies for different samples, which appear close for different labyrinthine patterns (except for the immediate vicinity of $g_{cr}$, where the fluctuations increase), and strongly differ from the energies of regular stripes. All points at the diagrams in **Figs.2** correspond to minimal energy.

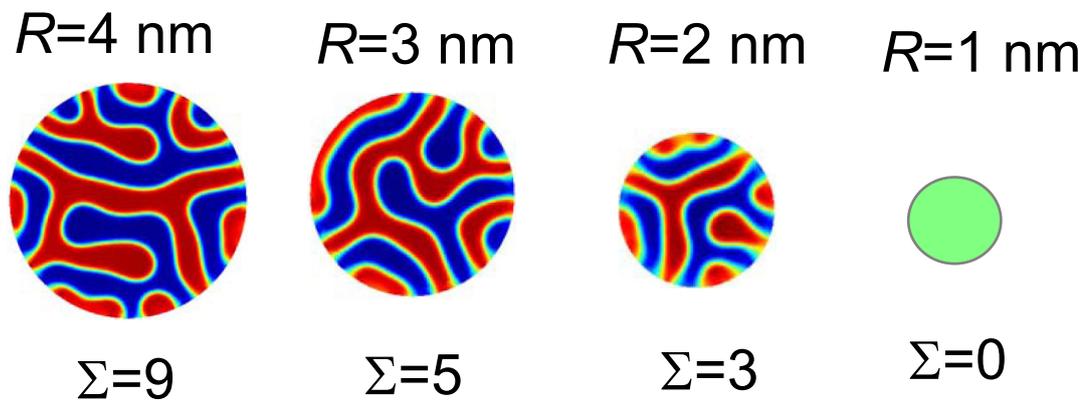

**FIG.S2.** Zoomed part of Fig.2(b).

**S.3. Clarifications concerning the charge of the domain walls**

Earlier thin ferroelectric films with nominally uncharged 180-degree domain structure have been considered and it has been shown that in the region of the domain wall contact with the electrically-open film surface the depolarization electric field appears and leads to the domain wall broadening near the surface [3, 4, 6]. The self-consistent calculations confirmed that broadening causes the decrease of depolarization field that in its turn leads to the decrease of the system electrostatic energy [3, 4, 6]. Notably, the broadened wall becomes curved and thus no more uncharged near the surface [3, 4, 6].

A similar effect was calculated by us for spherical particles and shown in the manuscript. Namely the domain walls in both PDFE and SLD phases are uncharged in the central part of the particle (far from its poles) and become charged and broadened at the polar surfaces [see **Fig. S3** and yellow-green regions near the poles in **Fig.1(a)**]. Far from the poles polarization vector (shown by white arrows in Fig.S5) is parallel to the wall planes and so the walls are uncharged. Inset shows the cross-section of the spatial region near the particle pole, where the polarization vector **P** in different domains (shown by red and blue arrows) is no more parallel to the curved surface of the broadened wall (walls are shown by black curves). The uncompensated bound charge $\sigma(\theta,\varphi)$ proportional to the projection of polarization normal $P_n(\theta,\varphi)$ on the domain wall surface occurs in the region of the broadened wall.

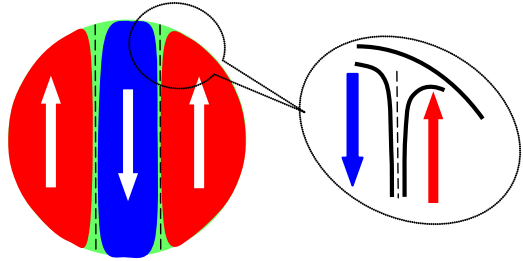

**FIG. S3.** Schematics of the domain wall broadening near the poles of the nanoparticle. Far from the poles polarization vector (shown by white arrows) is parallel to the wall planes and so the walls are uncharged. Inset shows the region near the particle pole, where polarization vector **P** in different domains (shown by red and blue arrows) are no more parallel to the curved surface of the broadened wall (shown by black curves).